# Effects of Navier Slip on Film Condensation Heat Transfer over Upward Facing Horizontal Flat Surfaces with Free Edges

Radi Alsulami[a*], Kannan Premnath[b], Mutabe Aljaghtham[c]

[a] *Department of Mechanical Engineering, Faculty of Engineering, King Abdulaziz University, Jeddah 21589, Saudi Arabia*

[b] *Department of Mechanical Engineering, College of Engineering, Design and Computing, University of Colorado Denver, Denver, CO 80217, U.S.A.*

[c] *Department of Mechanical Engineering, College of Engineering, Prince Sattam Bin Abdulaziz University, Al Kharj 11942, Saudi Arabia*

\* Corresponding Author: Radi Alsulami

Department of Mechanical Engineering, King Abdulaziz University, Jeddah, Saudi Arabia

Room No. 24E36, Building No. 40, Faculty of Engineering. P.O. Box 80204, Jeddah-21589, Saudi Arabia. Phone: +966548417334

E-mail: rasalsulami@kau.edu.sa

# Abstract

Transport phenomena involving condensate liquids generated from the phase change heat transfer in microchannels and in engineered superhydrophobic surfaces require consideration of slip effects. In this study, the laminar film condensation over upward facing flat slabs and circular disks of finite sizes with free edges in the presence of wall slip effects is investigated. By considering the Navier slip model and extending the classical Nusselt analysis, the mass, momentum, and energy of the liquid film in two-dimensional and axisymmetric coordinates are solved for the film thickness and the heat transfer rate in non-dimensional form. Numerical solution yields the local structure of the condensate film profile and the Nusselt number for different values of the slip coefficient. Investigation of the results reveals that the condensate film on horizontal surfaces becomes thinner and the overall heat transfer rate is enhanced with an increase in the slip coefficient. In particular, a regression analysis of the results indicates a power law dependence of the Nusselt number on the non-dimensional slip coefficient with an exponent close to 0.5. Significant enhancement in phase change heat transfer follow from the modification of the local velocity profiles within the condensate film, especially in resulting from the additional momentum gain near the wall surfaces due to increases in slip effects.

# Nomenclature

| Symbols | Description |
|---|---|

**Abbreviation**

| | |
|---|---|
| $c_p$ | specific heat at constant pressure |
| $D$ | diameter of horizontal disk Eq. (31) |
| $f$ | Factor |
| $f(x)$ | function Eq. (3) |
| $g$ | acceleration due to gravity |
| $h_f$ | enthalpy of a saturation liquid |
| $h_{fg}$ | latent heat of condensation, or vaporization |
| $h'_{fg}$ | modified latent heat of condensation, or vaporization |
| $h_g$ | enthalpy of saturated vapor |
| $\bar{h}$ | average heat transfer coefficient |
| $H$ | enthalpy flow rate Eq. (10) |
| $k$ | thermal conductivity |
| $L$ | width of horizontal plate |
| $\dot{m}$ | condensate mass flow rate |
| $\dot{m}(r)$ | condensation flow rate on disk of radius $r$ Eq. (25) |
| $\overline{Nu}_D$ | disk Nusselt number Eq. (18) |
| $\overline{Nu}_L$ | flat plate Nusselt number Eq. (34) |
| $p$ | pressure in the liquid |
| $p_{vap}$ | pressure in the vapor |
| $q'$ | total heat transfer rate into the slab, per unit length of slab Eq. (16) |
| $q''$ | surface heat flux |
| $r$ | radial position |
| $\hat{r}$ | dimensionless radial position, Eq. (30) |
| $T_{sat}$ | saturation temperature, and temperature of liquid-vapor interface |
| $T_w$ | temperature of solid surface |
| $u$ | liquid velocity |
| Y | a factor |
| W | a factor |
| Z | a factor |

**Greek Symbols**

| | |
|---|---|
| $\Gamma(x)$ | Condensation mass flow rate |
| $\beta_s$ | Navier slip coefficient |
| $\tilde{\beta}_s$ | dimensionless slip coefficient |
| $\delta$ | film thickness |
| $\tilde{\delta}$ | dimensionless film thickness |

| Symbol | Description |
|---|---|
| $\mu$ | dynamic viscosity |
| $\xi$ | dimensionless horizontal position |
| $\rho$ | Density |

## Subscripts

| *Symbols* | *Description* |
|---|---|
| $l$ | liquid |
| $V$ | vapor |

| *Symbols* | *Description* |
|---|---|
| $x, y, z$ | Cartesian coordinate directions |

# 1. Introduction

Condensation is a convection phase-change process which occurs when a saturated vapor converts to liquid by contact with the surface at a lower temperature [1]. The condensation process has several applications in industry and environment. As examples, it is one of the main phenomena in petroleum refinery, many distillation processes, steam generation equipment and thermal desalination systems [2]. The condensation phenomena is also significant in the design of heat exchangers [3] and modern manufacturing applications, such as mass soldering process (also called condensation soldering), which is a specific heat transfer method for reflow soldering and widely used in the electronics manufacturing [4–7]. Furthermore, the film condensation along a vertical flat plate under sinusoidal g-jitter in presence of the slip effect is important in microgravity space applications [8], while the film condensation on a horizontal plate in the presence of slip effect is useful in micro-flow in micro-electro mechanical systems (MEMS) applications [5,9].

Generally, many studies on film condensation heat transfer essentially comply with no-slip boundary conditions at the solid interface [10]. This is a result of the fact that for macroscopic flows of simple fluids, the slip effect/length is usually negligible, and the no-slip boundary condition can serve as effective assumption without loss of accuracy [11]. However, the effect of deviation from the traditional no-slip assumption becomes pronounced, specifically when the scale of interest reaches out micro and nano-flows [12], as well as when liquid flows on hydrophobic or superhydrophobic surfaces [13,14]. The use of hydrophobic and superhydrophobic surfaces is a result of their water repellency properties, which can aid in solving several engineering challenges, such as drag reduction, anti-icing and enhancement of two-phase heat transfer performance [13,15,16].

Most of the previous works have analytically and experimentally investigated the condensation heat transfer over a vertical plate, e.g., [17–22]. Al-Jarrah et al. [23] have investigated the effects of the velocity and temperature slip on film condensation in microchannels between two vertical plates. The authors showed that the consideration of slip length condition reflects on a non-zero slip velocity at the interface surface, leading to reduced condensate film thickness and boosted condensation mass flow rate per unit width. Similarly, Pati et al. [24] examined how the slip velocity can influence the condensation heat transfer over vertical surfaces by introducing a slip length condition, defined as the distance behind the solid interface at which the liquid velocity extrapolates to zero, into the classical condensation theory. They demonstrated that when the slip length increases, the condensate film thickness over the surface decreases, in comparison to classical theory (i.e., no-slip condition). This explained as an increase of slip value is associated with an augmentation of the slip velocity at the surface, resulting in a thinner condensate layer, and consequently enhancement in the heat transfer. In related work, Del Col et al. [10] have experimentally investigated the effect of hydrophobicity on the liquid film and the heat transfer coefficient of vertically mounted hydrophobic surfaces. They found that the heat transfer enhanced as the hydrophobicity (i.e., slip value) increased. This enhancement is believed to be caused by the reduction of the condensate film as a result of the augmentation of slip values for these surfaces.

Some works in the literature studied the condensation film flow in the presence of nonfluids. This is because nanofluids involve unique properties, including high thermal conductivity and low susceptibility to sedimentation, fouling, erosion and clogging [25]. For instance, Turkyilmazoglu [25] have extended the classical theory of Nusselt of regular film condensate by adding nanoparticles, such as Ag, Cu, CuO, Al2O3 and TiO2, while all fluid properties being dependent on the constant concentration of nanoparticles, that is the nanoparticles are uniformly and constantly distributed across the boundary layer within the single phase analysis. For this case, the author has analytically analyzed the laminar convective heat and mass transfer for a vertical surface with no-slip boundary condition and found that the presence of nanofluid enhances heat transfer rate by reducing the thickness of film condensate.

Few previous studies, however, have been performed on the condensation film flow over *horizontal* surfaces. Analytical studies on the film condensation on an upward facing plate with free edges were discussed by handful of authors. Bejan treated the upward facing horizontal plate with free edges with no slip effect [26]. Before that study, related problems were discussed by Rohsenow et al. [27] within a simpler setting under the assumption that the plate serves as the bottom surface for a vessel with adiabatic vertical walls and also by Gerstmann and Griffith [28] on the condensation on the underside of a horizontal plate. Chiou and Chang [4] studied a laminar film condensation on a horizontal disk with suction at the wall. They found that the dimensionless film thickness along the disk was a function of a parameter related to the ratio of Jakob number to Prandtl number and a suction parameter. Similar problem was discussed by Yue-Tzu et al. [29] by considering a finite- size horizontal wavy disk. In their study, the heat transfer coefficient and the film condensate thickness along the surface are found to be a function of the above mentioned parameters and the dimensionless wavelength.

To the best of the authors' knowledge, none of the previous works considered the effect of slip condition on film condensation on *horizontal* surfaces with free edges. In addition, due to the current interest in hydrophobic and superhydrophobic surfaces, as well as in microchannels, it becomes significantly important to study the effect of slip conditions on the heat transfer performance for different application and configurations. For the mentioned motivation, an appropriate physical model has been developed to study the effect of the wall slip on film condensation, condensate velocity, and heat transfer rate of two horizontal upward facing configurations, i.e., rectangular and circular plates with free edges. Wall slip effects of the condensate are represented using the Navier boundary condition, where the slip velocity is considered to be proportional the velocity gradient at the wall and parameterized by a coefficient often termed as the slip length. The classical Nusselt analysis is then extended by including such effects on horizontal condensation on upward facing flat surfaces in rectangular and axisymmetric configurations. Considering slip effects on the film condensation can provide a better understanding of the physics involved in terms of studying the influence of the slip coefficient on the horizontal wall condensation thickness and on the heat transfer rate. As will be shown in the next section, the presence of slip effect introduces additional terms in the governing equations of the dimensionless condensate film thickness. It may be noted that the resulting scaling exponents for the Nusselt number in terms of the various governing parameters is different for the horizontal surfaces compared to the conventional vertical surface condensation. Moreover, as a new result arising from this study, the Nusselt number for condensation heat transfer on horizontal surfaces is shown to depend on a numerical factor that depends on an appropriately non-dimensionalized magnitude of slip (i.e., Navier slip coefficient) in the form of a power law whose exponent is subsequently deduced from an analysis of the numerical solution.

## 2. Physical Model

A physical model for film condensation over a two dimensional and axisymmetric horizontal plate in the presence of slip effects is derived in this section.

## 2.1 Slip Effects on Film Condensation over a Horizontal Plate

Figure 1 illustrates the schematic of the condensate layer model representing a simplified formulation using an integral analysis is developed to obtain the expression for the velocity profile and the film thickness of the condensate layer. Then, a shooting method will be applied to solve the resulting film thickness equation numerically, which provides the spatial variation of the film thickness and heat transfer rate for different values of the slip coefficient in the Navier boundary condition.

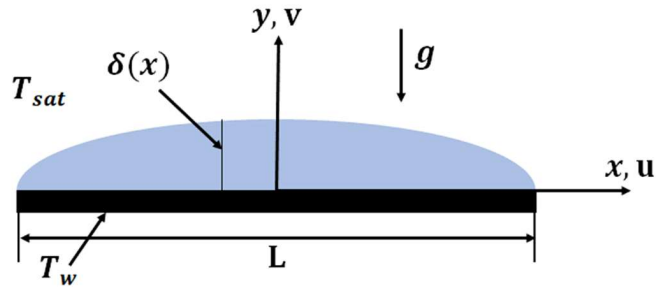

Figure 1. Schematic of condensate film on a horizontal slab.

In this case, the classical Nusselt-type analysis will be considered, except it will be on a *horizontal* plate maintained at relatively low-temperature $T_w$ facing up to the ambient at a saturated vapor temperature $T_{sat}$ as shown in Fig. 1. The rectangular plate is of width $L$ and unit depth normal to the paper. The momentum equations for the liquid (condensate) film layer are simplified based on the boundary layer approximation with negligible inertia effects, which, in the direction parallel and perpendicular to the plate, respectively are given as

$$0 = -\frac{\partial p}{\partial x} + \mu_l \frac{\partial^2 u}{\partial y^2}, \tag{1}$$

$$0 = -\frac{\partial p}{\partial y} - \rho_l g. \tag{2}$$

By integrating Eq. (2), we get

$$p(x, y) = -\rho_l gy + f(x). \tag{3}$$

Since the pressure distribution in the vapor region is hydrostatic and balances that with the liquid at the interface, the pressure at the edge of the condensate layer can be written as

$$p(x, \delta) = p_{vap}(\delta) = -\rho_l g\delta + f(x)$$

which leads to

$$p(x, y) = -\rho_l gy + p_{vap}(\delta) + \rho_l g\delta. \tag{4}$$

For the vapor phase, as mentioned above, the pressure distribution is hydrostatic in nature and can be written as

$$p_{vap}(y) = -\rho_v gy + \text{constant}.$$

Thus, it follows that the pressure field within the condensate layer may be finally written as

$$p(x, y) = (\rho_l - \rho_v)g\delta - \rho_l gy + \text{constant}. \tag{5}$$

By substituting Eq. (5) in Eq. (1), we get the simplified form of the momentum equation for the film condensate over a horizontal slab as the following

$$\mu_l \frac{\partial^2 u}{\partial y^2} = (\rho_l - \rho_v)g \frac{d\delta}{dx}. \tag{6}$$

The boundary conditions imposed for our physical model can be written as

$$u(y = 0) = \beta_s \frac{\partial u}{\partial y}\bigg|_{y=0}, \qquad \frac{\partial u}{\partial y}\bigg|_{y=\delta} = 0. \tag{7}(a,b)$$

In the boundary condition equation (7a), the slip velocity between the horizontal plate and the liquid film (condensate) is assumed to obey the Navier condition, which can model slip effects in liquids under a variety of conditions and configurations and various phenomenological mechanisms have been proposed for its interpretation [30-35]. Here, the factor $\beta_s$ is the slip length coefficient, which can be obtained experimentally and depend on the type of fluid and the nature of the surface. In Eq. (7b), zero shear is assumed at the liquid-vapor interface. By integrating Eq. (6) twice and applying the boundary conditions equations (7a) and (7b), we get the following equation for the velocity profile $u(x, y)$

$$u(x,y) = \frac{g}{\mu_l}(\rho_l - \rho_v)\frac{d\delta}{dx}\left(\frac{y^2}{2} - \delta y - \beta_s \delta\right). \tag{8}$$

In addition, the condensate flow rate per unit length normal to Fig. 1 obtained using Eq. (8) is given by

$$\Gamma(x) = \rho_l \int_0^\delta u(x,y) dy = -\frac{\rho_l g}{3\mu_l}(\rho_l - \rho_v)\frac{d\delta}{dx}[\delta^3 + 3\beta_s \delta^2]. \tag{9}$$

Clearly, the presence of slip enhances the condensate flow rate due to the presence of an additional term in Eq. (9) that depends on $\beta_s$, which, as discussed later, contributes towards enhancing the overall heat transfer rate. By invoking the usual assumption that the local temperature is distributed linearly across the condensate film (see e.g., [26]), the enthalpy flow rate is written as

$$H = \left[h_f - \frac{3}{8}c_{p,l}(T_{sat} - T_w)\right]\Gamma(x), \tag{10}$$

and the heat flux absorbed by the plate wall follows from applying the Fourier's law at the wall, which reads as

$$q''_w = \frac{k_l(T_{sat} - T_w)}{\delta}. \tag{11}$$

Following the classical Nusselt approach, the application of the energy balance for a differential control volume $\delta dx$ accounting for latent heat release during phase change (condensation) requires that in the steady state

$$0 = H - (H + dH) + h_g d\Gamma - q''_w dx, \tag{12}$$

where $h_g$ is the specific enthalpy of the vapor phase. By using Eqs. (9), (10) and (11) in the previous equation (Eq. (12)), we obtain the final differential form of the energy balance within the condensate as

$$\frac{k_l}{\delta}(T_{sat} - T_w)dx = h'_{fg} d\Gamma, \tag{13}$$

where $h'_{fg} = h_{fg} + \frac{3}{8} c_{p,l}(T_{sat} - T_w)$.

Then, by combining Eq. (13) with the expression for $\Gamma(x)$ given in Eq. (9) and simplifying, we get the following non-linear second-order ordinary differential equation for the condensate film thickness in non-dimensional form

$$-\tilde{\delta} \frac{d}{d\xi}\left(\tilde{\delta}^2 \frac{d\tilde{\delta}}{d\xi}[\tilde{\delta} + 3\tilde{\beta}_s]\right) = 3, \tag{14}$$

where the dimensionless variables used in the last equation are defined as

$$\xi = \frac{x}{L}, \quad \tilde{\delta} = \delta\left[\frac{h'_{fg}\rho_l(\rho_l-\rho_v)g}{k_l(T_{sat}-T_w)\mu_l L^2}\right]^{1/5}, \quad \tilde{\beta}_s = \beta_s\left[\frac{h'_{fg}\rho_l(\rho_l-\rho_v)g}{k_l(T_{sat}-T_w)\mu_l L^2}\right]^{1/5} \tag{15}$$

Here, $\tilde{\delta}$ and $\tilde{\beta}$ are the dimensionless condensate thickness and slip length coefficient, respectively. Equation (14) for the dimensionless condensate film thickness over a slip-compliant horizontal surface represents one of the new results of this study. The total heat transfer rate absorbed by the wall given in terms of that per unit length $q'$ as

$$q'L = 2\int_0^{L/2} k_l \frac{(T_{sat}-T_w)}{\delta} dx = 2k_l L(T_{sat} - T_w)\left[\frac{h'_{fg}\rho_l(\rho_l-\rho_v)g}{k_l(T_{sat}-T_w)\mu_l L^2}\right]^{1/5}\left(\int_0^{1/2} \frac{d\xi}{\tilde{\delta}}\right). \tag{16}$$

Now, the average heat transfer coefficient $\bar{h}$ is given by $\bar{h} = q'/(T_{sat} - T_w)$. Using Eq. (16), we can then deduce the expression for the heat transfer coefficient as

$$\bar{h} = 2k_l \left[\frac{h'_{fg}\rho_l(\rho_l-\rho_v)g}{k_l(T_{sat}-T_w)\mu_l L^2}\right]^{1/5} \left(\int_0^{1/2} \frac{d\xi}{\tilde{\delta}}\right). \tag{17}$$

Hence, the Nusselt number will then be given as the following equation

$$\overline{Nu_L} = \frac{\bar{h}L}{k_l} = \left(2\int_0^{1/2} \frac{d\xi}{\tilde{\delta}}\right)\left[\frac{L^3 h'_{fg}\rho_l(\rho_l-\rho_v)g}{k_l(T_{sat}-T_w)\mu_l}\right]^{1/5}. \tag{18}$$

Based on Eq. (18), we can identify a pre-factor $f = 2\int_0^{1/2} \frac{d\xi}{\tilde{\delta}}$, which will play an important role in the present analysis. This factor depends on the dimensionless condensate film thickness profile (via Eq. (14)). Notice the scaling exponent for the Nusselt number for the horizontal plate case is 1/5, which contrasts that for the standard vertical plate case, where it is 1/4. It is important observe here that the Nusselt number in Eq. (18) includes the factor term $f$ noted above which is parameterized by the dimensionless slip coefficient $\tilde{\beta}_s$ through the dimensionless liquid film thickness profile. In the special case involving the classical situation without the slip effects, i.e., $\tilde{\beta}_s = 0$, this pre-factor reduces to a constant value given as 1.079 (see e.g., [26])

Since the dimensionless variables $\xi$ and $\tilde{\delta}$ appear under an integral in the equation for the Nusselt number (see Eq. (18)), Eq. (14) will be first numerically solved by means of a shooting method. Once $\tilde{\delta}(\xi)$ is known, the values of the proportionality factor $f$ in the Nusselt number expression can then be determined for different values of the slip coefficient $\tilde{\beta}_s$ via a numerical quadrature.

## 2.2 Slip Effects on Film Condensation over a Horizontal Disk

A physical model of the film condensation over a horizontal disk in the presence of slip effects is derived in this section, where the velocity profile and the condensate film thickness are obtained from a mathematical formulation based on an integral analysis.

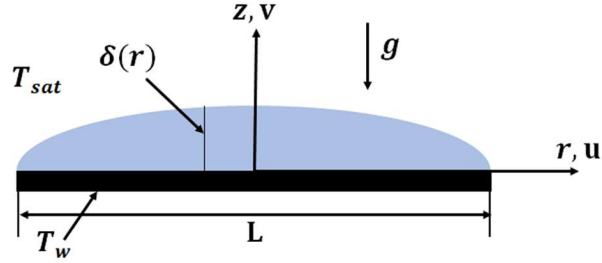

Figure 2. Schematic of condensate film on a horizontal disk.

Figure 2 illustrates a schematic of a condensate film on a horizontal disk of diameter $D$. Again, the Nusselt analysis will be applied on a horizontal disk facing up to the ambient maintained at relatively low-temperature $T_w$, similar to the horizontal rectangular plate case except that the axisymmetric cylindrical coordinates in the governing equations will be used for this configuration. Hence, the continuity and radial momentum equations are given, respectively, as

$$\frac{1}{r}\frac{\partial}{\partial r}(ru) = 0, \tag{19}$$

$$0 = -\frac{\partial p}{\partial r} + \mu_l \frac{\partial^2 u}{\partial z^2}. \tag{20}$$

Based on the axial momentum equation, the pressure distribution in the condensate layer can be written as

$$p(r,z) = (\rho_l - \rho_v)g\delta - \rho_l g z + \text{constant}. \tag{21}$$

By substituting Eq. (21) in Eq. (20) we get the reduced form of the momentum equation in the film condensate over a horizontal disk as follows:

$$\mu_l \frac{\partial^2 u}{\partial z^2} = (\rho_l - \rho_v) g \frac{d\delta}{dr}. \tag{22}$$

The boundary conditions for the present case are

$$u(z=0) = \beta_s \left.\frac{\partial u}{\partial z}\right|_{z=0}, \qquad \left.\frac{\partial u}{\partial z}\right|_{z=\delta} = 0, \tag{23)(a,b}$$

where, as before, the Navier slip condition is again applied at the wall of the disk (see Eq. (23a)).

By integrating Eq. (22) twice and applying the boundary conditions equations, the following equation for the velocity profile $u(r,z)$ is obtained:

$$u(r,z) = \frac{g}{\mu_l}(\rho_l - \rho_v)\frac{d\delta}{dr}\left(\frac{z^2}{2} - \delta z - \beta_s \delta\right). \tag{24}$$

Then, the radial condensate liquid flow rate through the film at radius $r$ follows as

$$\dot{m}(r) = 2\pi r \rho_l \int_0^\delta u(r,z)dz = -\frac{2\pi}{3\mu_l}\rho_l g (\rho_l - \rho_v)\frac{d\delta}{dr}[\delta^3 + 3\beta_s \delta^2]. \tag{25}$$

The heat flow rate by conduction at the wall over an elemental annular area is given in terms of the following equation involving the heat flux $q''$

$$q = \frac{k_l(T_{sat} - T_w)}{\delta}(2\pi r dr), \tag{26}$$

while differential mass flow rate across the condensate thickness over such an infinitesimal control volume is given by

$$d\Gamma = \frac{d}{dr}(\dot{m}(r))dr \tag{27}$$

Now applying Eq. (25) in Eq. (27) and performing the differentiation, we get the net condensate flow rate over the infinitesimal control volume in the presence of slip effects as

$$d\Gamma = -\frac{2\pi\rho_l g}{3\mu_l}(\rho_l - \rho_v)\left[r\delta^2(\delta + 3\beta_s)\frac{d^2\delta}{dr^2} + 3r\delta(\delta + 2\beta_s)\left(\frac{d\delta}{dr}\right)^2 + \delta^2(\delta + 3\beta_s)\frac{d\delta}{dr}\right]dr. \qquad (28)$$

Applying the energy conservation by accounting for the latent heat effects due to condensation over such an infinitesimal control volume, we get

$$q = h'_{fg} d\Gamma. \qquad (29)$$

Substituting Eqs. (26) and (28) in the previous equation (Eq. (29)) and simplifying, we get the following non-dimensional form of second non-linear ODE for the film thickness as

$$\hat{r}\tilde{\delta}^3(\tilde{\delta} + 3\tilde{\beta}_s)\frac{d^2\tilde{\delta}}{d\hat{r}^2} + 3\hat{r}\tilde{\delta}^2(\tilde{\delta} + 2\tilde{\beta}_s)\left(\frac{d\tilde{\delta}}{d\hat{r}}\right)^2 + \tilde{\delta}^3(\tilde{\delta} + 3\tilde{\beta}_s)\frac{d\tilde{\delta}}{d\hat{r}} + 3\hat{r} = 0, \qquad (30)$$

where the dimensionless variables used in the last equation are given as

$$\hat{r} = \frac{r}{D/2}, \quad \tilde{\delta} = \delta\left[\frac{h'_{fg}\rho_l(\rho_l - \rho_v)g}{k_l(T_{sat} - T_w)\mu_l D^2}\right]^{1/5} \text{ and } \tilde{\beta} = \beta\left[\frac{h'_{fg}\rho_l(\rho_l - \rho_v)g}{k_l(T_{sat} - T_w)\mu_l D^2}\right]^{1/5}. \qquad (31)$$

Clearly, the spatial distribution of the dimensionless condensate thickness $\tilde{\delta}$ is parameterized by the slip length coefficient $\tilde{\beta}_s$ as it appears in three of the terms in the condensate film thickness equation (Eq. (30)) and is a corresponding new theoretical result for the disk case.

Now, the total heat transfer rate absorbed by the wall of the circular disk of radius ($D/2$) is given as $\pi(D/2)^2 q'' = 2\pi \int_0^{D/2} k_l(T_{sat} - T_w)/\delta \, r dr$. After integrating the previous equation, we get the following expression for the heat flux $q''$:

$$\pi\left(\frac{D}{2}\right)^2 q'' = 2\pi k_l(D/2)^2(T_{sat} - T_w)\left[\frac{h'_{fg}\rho_l(\rho_l - \rho_v)g}{k_l(T_{sat} - T_w)\mu_l(D/2)^2}\right]^{1/5}\left(\int_0^1 \frac{\hat{r}}{\tilde{\delta}}d\hat{r}\right). \qquad (32)$$

Then, by using the definition of the heat transfer coefficient as $\bar{h} = q''/(T_{sat} - T_w)$ and substituting for $q''$ from Eq. (32), we obtain

$$\bar{h} = 2k_l \left[ \frac{h'_{fg}\rho_l(\rho_l-\rho_v)g}{k_l(T_{sat}-T_w)\mu_l D^2} \right]^{1/5} \left( \int_0^1 \frac{\hat{r}}{\tilde{\delta}} d\hat{r} \right). \tag{33}$$

Hence, from Eq. (33), we finally obtain the following expression for the Nusselt number for film condensation over an upward facing horizontal disk of diameter $D$ with a slip-compliant surface:

$$\overline{Nu}_D = \frac{\bar{h}D}{k_l} = \left( 2 \times 4^{1/5} \int_0^1 \frac{\hat{r}}{\tilde{\delta}} d\hat{r} \right) \left[ \frac{D^3 h'_{fg}\rho_l(\rho_l-\rho_v)g}{k_l(T_{sat}-T_w)\mu_l} \right]^{1/5}. \tag{34}$$

Here, we can then identify the pre-factor $f = 2 \times 4^{1/5} \int_0^1 \frac{\hat{r}}{\tilde{\delta}} d\hat{r}$, appearing in Eq. (34) for the Nusselt number. The sensitivity of the slip coefficient $\tilde{\beta}_s$ on the Nusselt number appears through this factor via the quadrature involving $\tilde{\delta}$, which depend on $\tilde{\beta}_s$ (See Eq. (30)).

## 3. Results and Discussion

### 3.1 Slip Effects on Film Condensation over a Horizontal Rectangular Plate

The nonlinear second-order differential equation for the film condensate thickness on a horizontal rectangular plate with free edges in the presence of the slip boundary condition (Eq. (14)) is solved using the shooting method for different values of the slip coefficient. The results show the impact of having the slip condition on the film condensation over a horizontal plate (e.g., with for hydrophobic or superhydrophobic horizontal surfaces). Figure 3 illustrates the variation of the dimensionless thickness $\tilde{\delta}$ with the dimensionless horizontal position $\xi$ for different values of the slip coefficient parameter $\tilde{\beta}_s$. This figure shows that an increase in the values of the slip coefficient $\tilde{\beta}_s$ (i.e., increase in the hydrophobicity property) results in a decrease in the thickness of the film condensate. he profile of the dimensionless thickness $\tilde{\delta}$ for the special case when the slip effect is not considered (i.e., $\tilde{\beta}_s = 0$) is quantitatively consistent with that provided in [26] for the no-slip case. The condensate has the largest thickness at the middle of the plate where $\xi = 0$. The reduction in the film thickness with $\tilde{\beta}_s$ is a manifestation of increased local fluid transport near the wall surface with a corresponding reduction in the local velocity near the liquid-vapor interface.

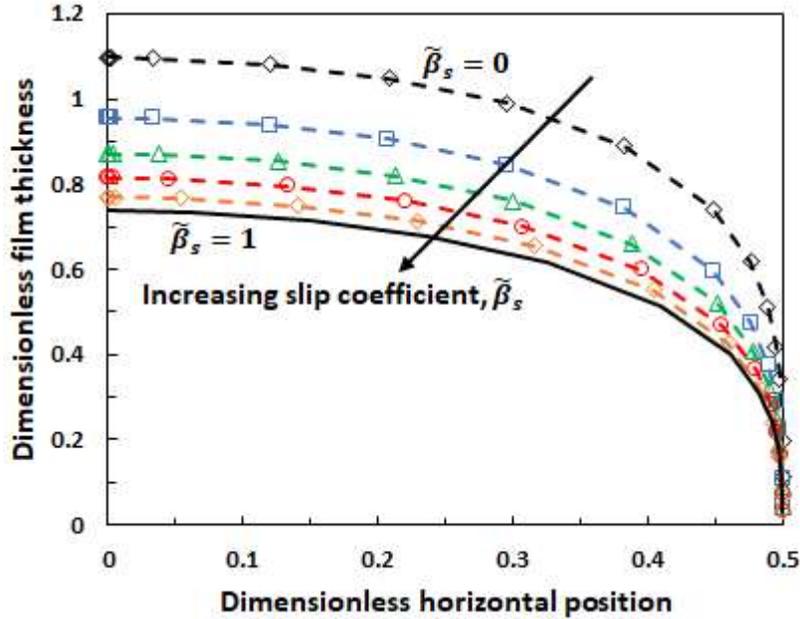

Figure 3. Variation of the condensate film thickness for different values of the slip coefficient $\tilde{\beta}_s$ (i.e., 0. 0.2. 0.4, 0.6, 0.8, and 1) over a horizontal upward facing rectangular plate.

Figure 4 shows the relationship between the slip coefficients $\tilde{\beta}_s$ and a scaled form of the Nusselt number given in terms of the pre-factor $f$, i.e., $\overline{Nu_L} / \left[ \frac{L^3 h'_{fg} \rho_l (\rho_l - \rho_v) g}{k_l (T_{sat} - T_w) \mu_l} \right]^{1/5}$. It clearly illustrates that an increase in slip coefficient results in a corresponding increase in the Nusselt number ($\overline{Nu_L}$). Thus, greater the magnitude of the condensate wall slip, larger is the film condensation heat transfer rate to the wall surface. This is a consequence of the fact an increase in the wall slip corresponds to augmentation of the condensate flow momentum near walls, which in turn leads to more pronounced convective thermal transport to the wall. Such enhancement of heat transfer rate at the wall arising from phase change (condensation) over slip-compliant surfaces is consistent with other studies involving increased flow and/or thermal transport over walls obeying the Navier slip condition (see e.g., [36]-[43]). Moreover, a regression curve fit to Fig. 4 yields the following best fit power law-type equation for the scaled Nusselt number dependence on the dimensionless slip coefficient $\tilde{\beta}_s$:

$$\frac{\overline{Nu_L}}{\left[\frac{L^3 h'_{fg} \rho_l (\rho_l - \rho_v) g}{k_l (T_{sat} - T_w) \mu_l}\right]^{1/5}} = c + d\tilde{\beta}_s^n, \tag{35}$$

where the constants $c$, $d$ and $n$ are given by 1.053, 0.676, and 0.4987, respectively. Thus, the Nusselt number appears to scale with the slip coefficient as $\overline{Nu_L} \sim \sqrt{\tilde{\beta}_s}$.

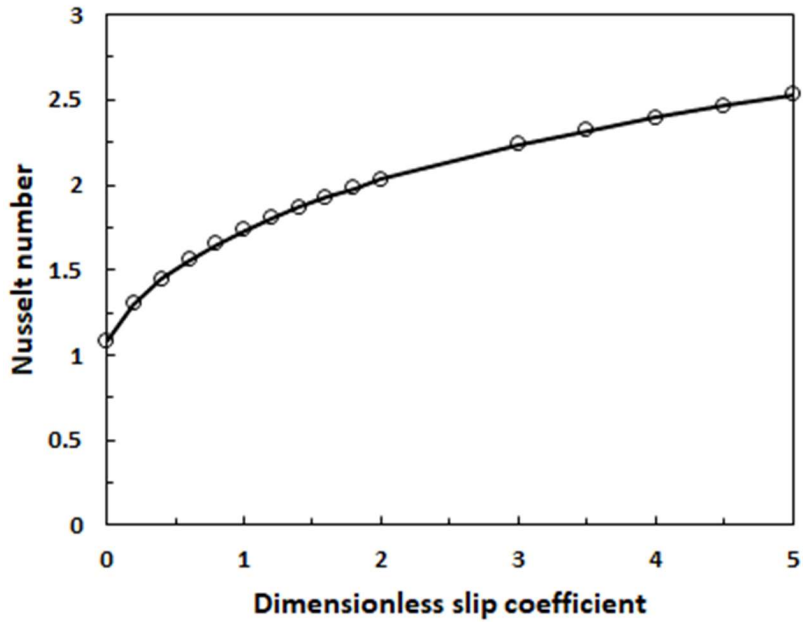

Figure 4. Effect of the slip coefficient $\tilde{\beta}_s$ on the heat transfer rate given in terms of the scaled Nusselt number in the case of condensate over a horizontal rectangular plate.

In order to further understand how the variations in the local flow structure within the condensate film with the slip coefficient contribute to the enhanced thermal transport during condensation mentioned above, we now plot the dimensionless velocity profiles for various $\tilde{\beta}_s$ at different locations within the film condensate. Thus, nondimensionalizing the velocity profile equation utilizing the following variables

$$\xi = \frac{x}{L}, \quad \tilde{\delta} = \frac{\delta}{W}, \quad \tilde{\beta}_s = \frac{\beta_s}{W}, \text{ and } \tilde{y} = \frac{y}{W}$$

where $W = \left[\frac{k_l(T_{sat}-T_w)\mu_l L^2}{h'_{fg}\rho_l(\rho_l-\rho_v)g}\right]^{1/5}$, the dimensionless velocity profile equation becomes

$$\tilde{u}(\xi, \tilde{y}) = \tilde{\delta}^2 \frac{d\tilde{\delta}}{d\xi}\left[\frac{1}{2}\left(\frac{\tilde{y}}{\tilde{\delta}}\right)^2 - \left(\frac{\tilde{y}}{\tilde{\delta}}\right) - \left(\frac{\tilde{\beta}_s}{\tilde{\delta}}\right)\right]$$

where $0 \leq \xi \leq 0.5$ to and $0 \leq \tilde{y} \leq \tilde{\delta}$.

The velocity profiles across the condensate thickness at three different locations from the center of the plate ($\xi = 0.1, 0.2, 0.3$) for various values of the dimensionless slip coefficient $\tilde{\beta}_s$ are shown in Figs. 5, 6 and 7.

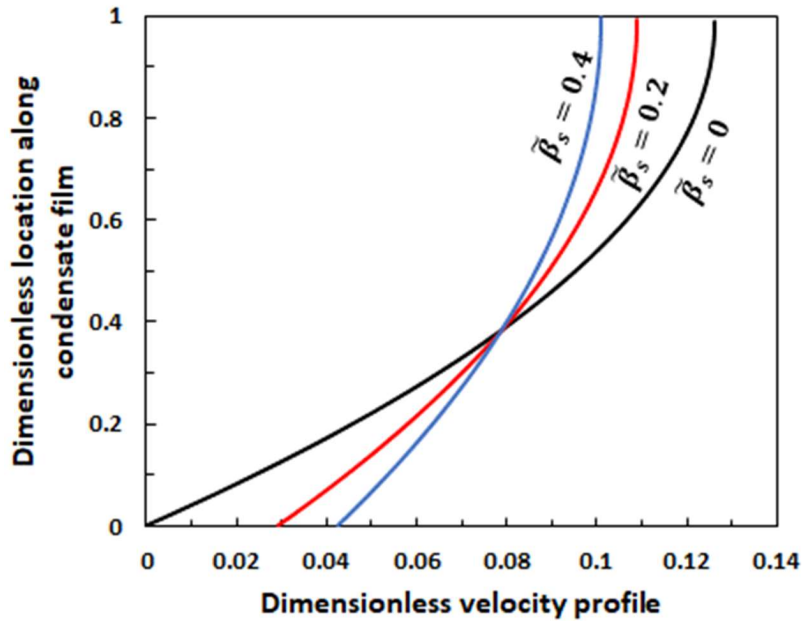

Figure 5. Variation of the velocity profile within the condensate film layer over a rectangular plate for different values of dimensionless slip coefficient $\tilde{\beta}_s$ at $\xi = 0.1$.

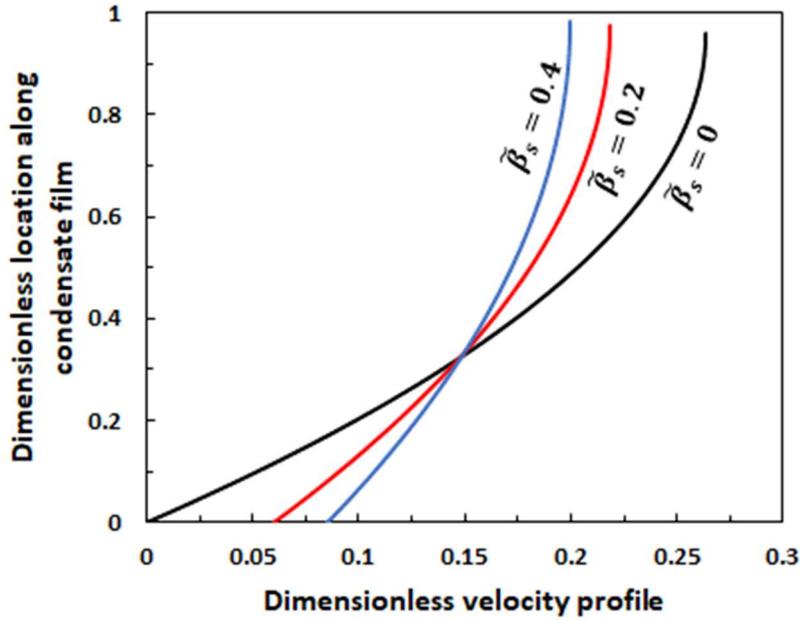

Figure 6. Variation of the velocity profile within the condensate film layer over a rectangular plate for different values of dimensionless slip coefficient $\tilde{\beta}_s$ at $\xi = 0.2$.

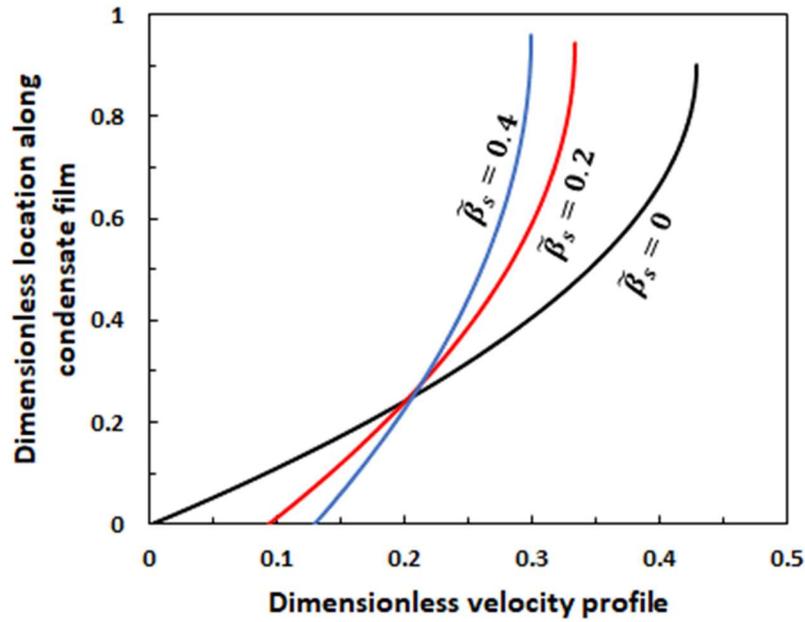

Figure 7. Variation of the velocity profile within the condensate film layer over a rectangular plate for different values of dimensionless slip coefficient $\tilde{\beta}_s$ at $\xi = 0.3$.

As the dimensionless slip coefficient $\tilde{\beta}_s$ increases, as expected, the near wall velocities within the condensate layer increases. Thus, in general, it is noticed that for a given location from the center, the momentum of the condensate is increased in the near wall region, thereby promoting the heat transfer rate as discussed above. Also, it is noticed that the magnitudes of the velocity in the region near the edge of the condensate are reduced with $\tilde{\beta}_s$ to maintain the overall mass conservation. Furthermore, by comparing Figs. 5, 6 and 7, it is observed that the farther the location from the center, the greater is the magnitude of the velocities for any fixed $\tilde{\beta}_s$.

**3.2 Slip Effects on Film Condensation over a Horizontal Circular Disk**

In this section, the film condensation over a horizontal circular disk in the presence of slip is studied. Figure 8 illustrates the variation of the dimensionless film thickness $\tilde{\delta}$ as a function of the dimensionless radial coordinate $\hat{r}$, at different values of the slip coefficient parameter $\tilde{\beta}_s$ (i.e., variation in hydrophobicity property), which is obtained by solving Eq. (30). Similar to the plate configuration discussed in the previous section, it is found that an increase in the values of the slip coefficient $\tilde{\beta}_s$ results in a decrease in the thickness of the film condensation similar to that for the horizontal plate case. Note that again the largest thickness of film condensate is in the middle of the disk where $\hat{r} = 0$. Also, the structure of the dimensionless thickness $\tilde{\delta}$ profile for the special case when slip effect is not considered (i.e., $\tilde{\beta}_s = 0$) is in good agreement with that presented in [26] for the no-slip case. It may be noted that comparing Figs. 3 and 8, the film condensate over a horizontal disk appears to be thicker at the centerline than that over a horizontal rectangular plate as a consequence of the geometric effect related to the differences in the growth of the surface area with the distance from the center between the two cases.

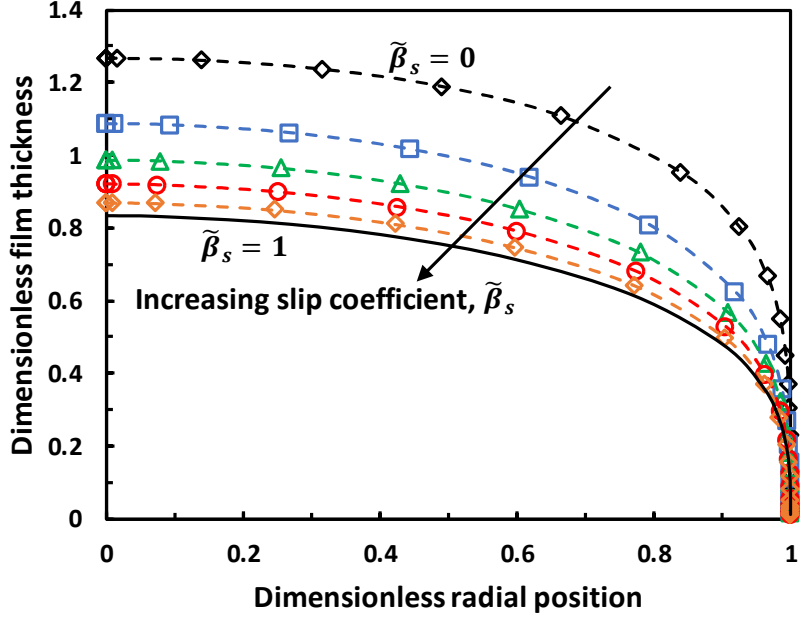

Figure 8. Variation of the condensate film thickness for different values of slip coefficient $\tilde{\beta}_s$ (i.e., 0. 0.2. 0.4, 0.6, 0.8, and 1) over a horizontal upward facing circular disk.

Figure 9 shows the relationship between the slip coefficients $\tilde{\beta}_s$ and the Nusselt number $\overline{Nu}_D$ in terms of the scaled Nusselt number, i.e., $\overline{Nu}_D / \left[\dfrac{D^3 h'_{fg} \rho_l (\rho_l - \rho_v) g}{k_l (T_{sat} - T_w) \mu_l}\right]^{1/5}$. Again, it clearly illustrates that an increase in the slip coefficient results in an increase in the Nusselt number ($\overline{Nu}_D$) reflecting an overall increase in the heat transfer rate. Also notice that the heat transfer rate in a horizontal disk due to film condensation is larger than that in a horizontal rectangular plate because of the fact that for the plate, the film condensate is thicker than that on a horizontal disk of the same width as $D$ where for the latter, the area progressively increases with the radial distance from the center. Finally, a regression curve fit to the numerical results in Fig. 9 yields the following power-law type equation expressing the parametric dependence of the scaled Nusselt number on the dimensionless slip coefficient:

$$\frac{\overline{Nu}_D}{\left[\dfrac{D^3 h'_{fg} \rho_l (\rho_l - \rho_v) g}{k_l (T_{sat} - T_w) \mu_l}\right]^{1/5}} = c + d\tilde{\beta}_s^n, \qquad (36)$$

where $c$, $d$ and $n$ are given by 1.323, 1.043, and 0.4662, respectively.

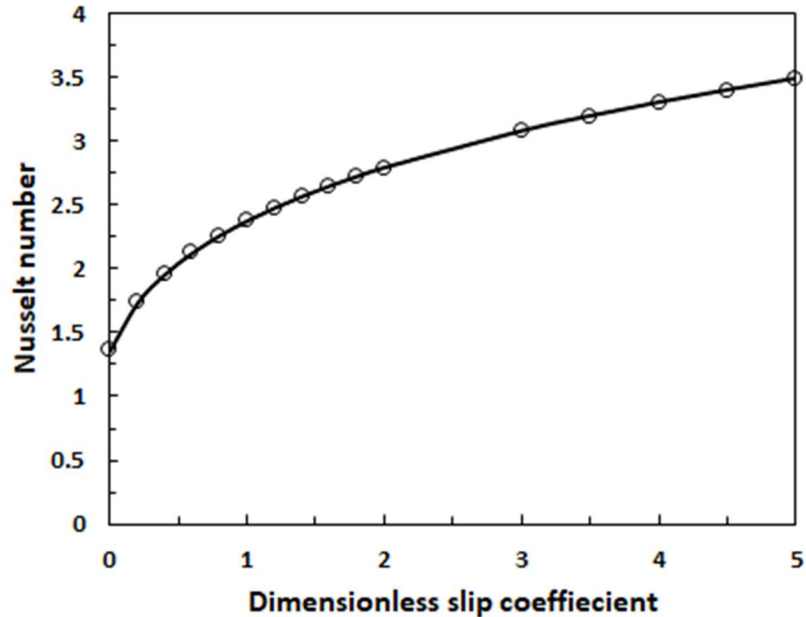

Figure 9. Effect of the slip coefficient $\tilde{\beta}_s$ on the heat transfer rate given in terms of the scaled Nusselt number in the case of condensation over a horizontal disk.

## 4. Conclusions

The problem of film condensation over an upward facing horizontal rectangular plate and a circular disk involving slip-compliant surfaces was formulated and studied. By making use of the standard Nusselt analysis along with the Navier slip condition at the wall surface, mathematical formulations of the physical models were developed for steady, laminar condensate flow and heat transfer over slipping wall surfaces in planar and axisymmetric coordinates. The resulting non-linear second-order differential equation for the structure of the dimensionless film thickness which now contains additional terms involving a dimensionless slip coefficient parameter $\tilde{\beta}_s$ were solved numerically using the shooting method.

It was shown that the film thickness of the condensate over both the horizontal rectangular plate and disk decreases as the slip coefficient increases. It was also found that an increase in the values of the slip coefficient results in an increase in the heat transfer rate. This was interpreted to be due to the increased momentum flux in the condensate near walls, when a velocity slip is present as compared to the no-slip case. In addition, it was noticed that the heat transfer rate in a horizontal disk is larger than in a horizontal plate due to the axial symmetry effects present in the former case accompanied by a thicker film condensate. In general, the normalized Nusselt number is found to vary with the slip coefficient $\tilde{\beta}_s$ via a power-law type equation as $c + d\tilde{\beta}_s^n$, where $c$, $d$ and $n$ are constants. In particular, regression fits to the data from the numerical solution yield the exponent $n$ as 0.498 and 0.466 for the plate and disk, respectively, which provides an approximate scaling relation as $\overline{Nu}_L \sim \tilde{\beta}_s^{1/2}$. Equivalently, as a key finding of this study, if $\overline{Nu}_{L,0}$ and $\overline{Nu}_{L,\beta}$ represent the Nusselt numbers reflecting the heat transfer rates during film condensation over horizontal surfaces without and with slip effects, respectively, then the relative heat transfer enhancement with using slip-compliant surfaces can be correlated with the slip coefficient parameter as

$$\frac{\overline{Nu}_{L,\beta} - \overline{Nu}_{L,0}}{\overline{Nu}_{L,0}} = \tilde{\beta}_s^n = \left\{ \beta_s \left[ \frac{h'_{fg} \rho_l (\rho_l - \rho_v) g}{k_l (T_{sat} - T_w) \mu_l L^2} \right]^{1/5} \right\}^n,$$

where $n \sim 0.5$. These results may serve as guidance in the design, analysis and interpretation of film condensation over special engineered surfaces, such as with superhydrophobicity, and in microchannel configurations.

## Acknowledgement

The first author gratefully acknowledges the support of this research by King Abdulaziz University, Jeddah, Saudi Arabia.